\documentclass[intlimits,twoside,a4paper]{article}

\usepackage{amsmath,amssymb}
\usepackage{graphicx,psfrag}
\usepackage[official]{eurosym}

\usepackage[cp1251]{inputenc}

\usepackage[eqsecnum]{cmpj3}
\usepackage{xcolor}
\issue{2020}{23}{2}{23002}
\doinumber{10.5488/CMP.23.23002}
\title[Two cultures]%
{Two cultures: ``them and us'' in the academic world}
\author[R. Kenna]{R. Kenna\refaddr{label4,label3}}
\addresses{
\addr{label4}Statistical Physics Group,
Centre for Fluid and Complex Systems, Coventry
University, Coventry, England \addr{label3}Doctoral
College for the Statistical Physics of Complex Systems,
Leipzig-Lorraine-Lviv-Coventry $({\mathbb L}^4)$}

\date{Received March 2, 2020, in final form April 11, 2020}

\begin{document}
\maketitle
\begin{abstract}
Impact of academic research onto the non-academic world is of increasing importance as authorities seek  {return on} public investment.
Impact opens new opportunities for {what are known as} ``professional services'':
as scientometrical tools {bestow} some with confidence they can quantify quality, the impact agenda brings lay measurements to evaluation of research.
This paper is partly inspired by the famous ``two cultures'' discussion  instigated by C.P. Snow over 60 years ago.
He saw a chasm between different academic disciplines and I see a chasm between academics and professional services, {bound} {{into contact through competing targets.}}
This paper draws on my personal experience and experiences recounted to me by colleagues in different
universities in the UK.
It is aimed at {igniting} discussions amongst people {interested in improving 
the academic world}
and it is intended in a spirit of collaboration and constructiveness.
As a professional services colleague said, what I have to say ``needs to be said''. 
It is my pleasure to submit this paper to the Festschrift devoted to the 60th birthday of a renowned physicist, my good friend and colleague Ihor Mryglod. 
Ihor's role as leader of the Institute for Condensed Matter Physics in Lviv has been essential to generating some of the impact described in this paper and forms a key element of the story I wish to tell.
\end{abstract}

%%%%%%%%%%%%%%%%%%%%%%%%%%%%%%%%%%%%%%%%%%%%%%%%%%%%%%%%%%%%%%%5
\section{Introduction}
\label{1}
%%%%%%%%%%%%%%%%%%%%%%%%%%%%%%%%%%%%%%%%%%%%%%%%%%%%%%%%%%%%%%%5

At the auspicious 1959  Rede Lectures in Cambridge University, C.P. Snow famously lamented what 
he perceived as a ``gulf of mutual incomprehension'' between ``literary intellectuals'' and scientists \cite{Snow59}. 
He worried that ``the intellectual life of the whole of western society'' was split into two polar groups and this chasm was a major hindrance to solving the world's problems.
His worriment was encapsulated in the  term ``two cultures'' and impacted massively on the spheres of academia and politics. 
This impact continues today --- at his opening address at the Munich Security Conference in 2014,  Estonian president Toomas Hendrik Ilves quoted Snow's famous essay when he said~\cite{Estonia}.
\begin{quote}
I think much of the problem we face today is the absence of dialogue between the scientific-technological and the humanist traditions. Bereft of understanding of fundamental issues and writings in the development of liberal democracy, computer geeks devise ever better ways to track people simply because they can and it's cool. Humanists on the other hand do not understand the underlying technology and are convinced, for example, that tracking meta-data means the government reads their emails.
\end{quote}

On this occasion of the 60's birthday of an  outstanding academic leader of another auspicious institution, I believe our {(academic)} world is faced with another problem and, again it involves two cultures. 
I perceive a ``gulf of mutual incomprehension'' between what are called ``professional services'' and academics, and I wish to instigate a ``two cultures'' discussion.
My worry {extends to} the intellectual life of society as academic credence
diminishes with the rise of a new layer of managers seemingly intent on corporatisation of universities at least here in the UK.

There are other parallels between Snow's worries and my own. Besides ``mutual incomprehension'' Snow perceived ``hostility and dislike, but most of all lack of understanding'' between the two cultures. ``They have a curious distorted image of each other. Their attitudes are so different that, even on the level of emotion, they can't find much common ground.''
Snow's description  matches {mutual perceptions} between many academics and many professional services today, forced into contact with each other as the corporate takeover of academia continues.

Literary intellectuals, Snow remarked, ``while no one was looking took to referring to themselves as `intellectuals' as though there were no others''. 
{The self-appellation ``professional services'' may similarly be interpreted, at least in some quarters, as suggesting there are no others~\cite{GibbsCU}.} 
I wish to discuss the consequences of this and the diminution of academic value, confidence and freedom.

%\begin{figure}[!b]
%\centerline{\includegraphics[angle=0, width=7.5cm]{experiencedscientists2b.eps}}
%\caption{(Colour online) 
%``{\emph{We are experienced scientists!}}'' --- a desperate plea for respect by an academic or a trophy on the wall of a ``professional services'' office? 
%\label{fig1}}
%\end{figure}

In giving my own personal experience of  the two cultures dichotomy  I do not wish to imply that never the twain shall meet.
Indeed, Snow describes
\begin{quote}
A good many times I have been present at gatherings of people who, by the standards of the traditional culture, are thought highly educated and who have with considerable gusto been expressing their incredulity at the illiteracy of scientists. Once or twice I have been provoked and have asked the company how many of them could describe the Second Law of Thermodynamics. The response was cold: it was also negative. Yet I was asking something which is the scientific equivalent of: Have you read a work of Shakespeare's?
\end{quote}
Many a time I have enjoyed convivial company of non-academic staff. 
Indeed I ended up marrying one of them! (Admittedly she did  convert to an academic in due course~\cite{Claire,Claire1}.)
But rarely have I quizzed them about the  second law of thermodynamics. 
I prefer to educate and to learn rather than to challenge my friends and, in the hope that this contribution will be read beyond exalted heights of the ``ivory towers'', I take the opportunity to briefly revise the laws of thermodynamics for those not versed in their succinctness. (I return to these laws in section~\ref{4.2} where I attempt to form analogously simple ``rules'' of {\emph{impact}}.)

The first law is that of conservation of energy: work and heat convert to the internal energy of a system.
The second law is that  entropy increases so that time moves forward. 
The third law is simply that entropy
approaches zero at zero temperature for a perfect system.
The zeroth law was named such because, although formulated after the first three, it is more fundamental. 
It
states that if a system is in   equilibrium with a second, and the second with a third, then the first is also in equilibrium with the third. 

So this is thermodynamics in a single paragraph. It is a ``top-down'' theory in that it doesn't offer where the ``laws'' actually come from. 
Instead they come  from ``on high'', as if to say ``this is the way the world is and that's that''. 
Their power is persuasive if not illuminative, so much so that in 1910, Henry Adams proposed adapting the second law of thermodynamics to develop a ``theory of history''~\cite{Adams1910}. 
In essence his idea was that, as entropy increases, order becomes disorder and  our world will ultimately become uninhabitable.

Physicists have long moved on from trying to apply top-down thermodynamics to society. 
Instead more emergent concepts have arisen from the  fundamental (``bottom-up'') discipline of statistical physics. 
For example, Ihor Yukhnovskii, who founded the Institute for Condensed Matter Physics 
(ICMP) in 1969,  is well known for works on applications of methods of theoretical physics to economy and society.
Complexity science and sociophysics are very much on the rise and the Lviv group remains to the fore in the emergence of these new disciplines.

It strikes me from my experience here in the UK, and accounts from many academic colleagues,  that, as the academic rules  are increasingly dished out from ``on high'', 
perhaps thermodynamic concepts may apply after all --- at least to see where they take us. 
Many perceive that everything is getting worse and  our academic worlds are becoming unbearable.
As one esteemed colleague described it, ``command and control'' is diminishing our academic freedom and ``corporate overlords''  are taking over. 
Is there a way to halt the tide of academic misery or at least find common ground? 
And if thermodynamics can't help us can statistical physics come to the rescue?

``{\emph{Impact}}'' is a relatively new key-word in academia~\cite{Martine}. 
In funding applications we now not only have to predict the academic success of our work but also the impact it will make beyond academia.
The theme of this miscellany is the latter type of impact (impact on the non-academic sectors beyond  academic journals, etc.).
It is society that funds our adventures up the ivory towers in the first place and the impact agenda is so that it receives something in return. 
Indeed, were it not for funding from governments (hence society) we may revert to the Victorian era where research was the preserve of the wealthy and people like me would not be writing papers like this (some may prefer that of course).
Economic impact on industry is often predicted as a promise ---  take a punt on our research you might profit in future and progress the human condition.  
Societal impact is another thing --- and can be hard (sometimes impossible) to predict. 
 
UK Research and Innovation (UKRI)  now considers impact   ``a core consideration throughout the grant application process''~\cite{core} and the European Commission has launched its Horizon Impact Award to recognise its most impactful projects~\cite{H2020impact}.
Here in Coventry, our mathematical sciences activities are mainly fluid dynamics and statistical physics and, as it happens, the second strand is the one that has generated most of our impact to date. A significant portion of that impact has emerged from collaboration with the ICMP in Lviv.
Ihor Mryglod is Director of that esteemed institution and it is my pleasure to dedicate this paper to him and the community he leads on this, the occasion of his 60th birthday.

The {Coventry-Lviv partnership} is based firmly on trust founded on  mutual academic respect --- we each have our strengths and together we are greater than the sum of our parts, {having} produced hundreds of papers since the foundation of the statistical physics group in Coventry (2007).
As trust between academics on either side of Europe has led to world leading academic impact, 
so too does non-academic impact require bridging the cultural gap between academia and professional services.
Since this notion is relatively new, this type of trust is far less developed.
\begin{figure}[!b]
	\centerline{\includegraphics[angle=0, width=7.5cm]{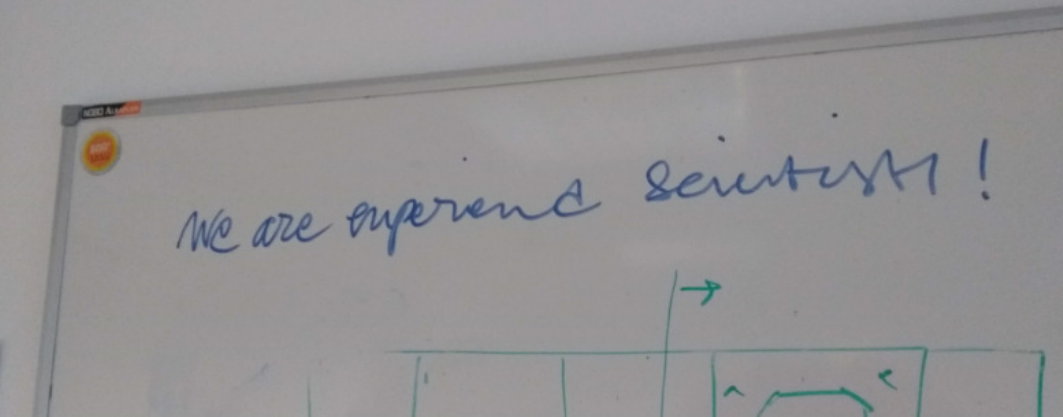}}
	\caption{(Colour online) 
		``{\emph{We are experienced scientists!}}'' --- a desperate plea for respect by an academic or a trophy on the wall of a ``professional services'' office? 
		\label{fig1}}
\end{figure}

While  {acknowledgment of} {\emph{impact}} as an intrinsic feature in research {may indeed be welcome}, I would like to take this opportunity to point out some of the travails it has instigated.
It forces together two cultures which have very different priorities.
Metrics are frequently developed by academics and deployed by professional services.
Ironically they are distrusted by the former and {more} trusted  by the latter.
The increased corporatisation of academia means that ``top-down, command-and-control''  structures
force professional services to chase academics for metrics, deadlines, exam uploads and paperwork for teaching, research and impact. 
Academics have targets to deliver and professional services have targets to ensure delivery.
The resulting tensions  need to be discussed and I see this paper
%the impact agenda 
as an opportunity to do so.
 
In section~\ref{2} I recount how the national research policy impacted on the academic landscape in the UK and how   Coventry's mathematical sciences team was able to navigate its way through it.
Using the insights and tools of statistical physics we were able to grow, thrive and adapt to each and every circumstance and even impact upon national and international research policies.
In section~\ref{3} I summarise the critical role played by academic trust in this journey and outline some of the non-academic impact that emanated from it.
In section~\ref{4} I give examples of debate between 
those {(mostly professional services)} with austere, linear views of impact and {those (mostly academics)} with broader views. 
{Drawing further from C.P. Snow's reference to the four laws of thermodynamics,}
I give four analogous
{succinct descriptors of impact.}
In section~\ref{5} I discuss the catastrophic consequences of culture-clash and in section~\ref{6}  I chart a path for the future of  mathematical sciences
at Coventry and perhaps the entire  academic sector.

{Sections~\ref{2} and~\ref{3} are retrospective and may be skipped by pressed readers.
Later sections are increasingly  prospective.
All views are my own or recounted to me in a number of universities 
and I use the liberal tradition of Festschriften to air them.}
Messages I wish to deliver through this paper include: don't blindly trust metrics 
{{and}} don't rely on individual opinions --- calibrate correctly;  just because you have a yardstick, don't insist on 
measuring the unmeasurable --- it just generates misery; don't be misled by a false sense of understanding --- be humble.
In other words, read figure~\ref{fig1} and trust us --- ``we are experienced scientists''!

%%%%%%%%%%%%%%%%%%%%%%%%%%%%%%%%%%%%%%%%%%%%%%%%%%%%%%%%%%%%%%%5
\section{Counting what counts in Coventry}
\label{2}
%%%%%%%%%%%%%%%%%%%%%%%%%%%%%%%%%%%%%%%%%%%%%%%%%%%%%%%%%%%%%%%5

The Further and Higher Education Act 1992  made sweeping changes to higher education in the UK. 
Thirty-five polytechnics were allowed to become universities and new funding bodies were formed.
The Higher Education Funding Council for England (HEFCE) was the largest of these until 2018, when it was transformed into teaching and research wings. 
The latter is Research England and operates within the UKRI.

% % % % % % % % % % % % % % % % % % % % % % % % % % % % % % % % % 
\subsection{Accountability of academia}
\label{2.1}
% % % % % % % % % % % % % % % % % % % % % % % % % % % % % % % % % 

{The Haldane Principle is the idea that researchers themselves are better able to judge the importance of research than bureaucrats. The principle dates from 1904 and still features  in UK  policy.}
The Research Assessment Exercise (RAE) was undertaken  every few years on behalf HEFCE and other UK funding councils up to 2008, after which it was replaced by the Research Excellence Framework (REF).
Both were established to evaluate the quality of research in all disciplines, categorised into Units of Assessment (UOA). 
Each UOA was/is assessed by a peer-review panel and the results used to determine funding for years to come. 
For example, in 2019--2020  \pounds{1.7} {billion was distributed.}
But the importance of REF extends beyond this for it determines  rankings and hence prestige of  universities, and these in turn influence student-recruitment numbers.
Annual tuition fees for ``home students'' are now close to  \pounds 10,000 and for overseas students can be far higher --- so it is indeed big business. 

\begin{figure}[!t]
\centerline{\includegraphics[angle=0, width=5.5cm]{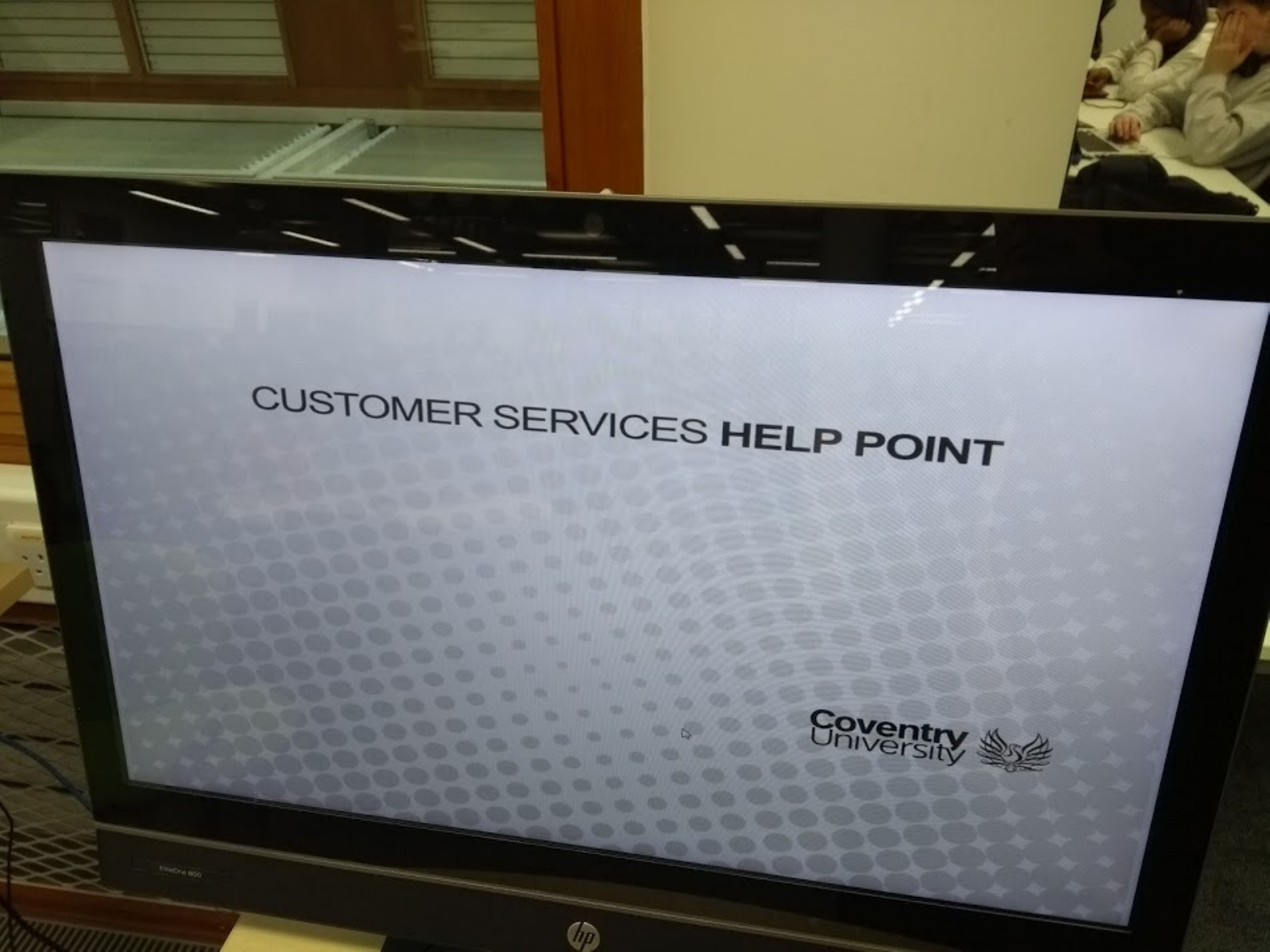}}
\caption{(Colour online) Image from Coventry-University Library illustrates the shifting mindshift --- what once were ``students'' are now referred to as ``customers''. 
 \label{fig2}}
\end{figure}

The mindshift to business is illustrated in figure~\ref{fig2} which displays an image from a university library referring to ``customers'' rather than ``students''. 
The shift is pervasive and similar photos can be taken all around the UK. 
The venerable concept of a ``collegium'' as a club or society of people
with the same goal of learning has been (and is being) eroded. 
This impacts on students too as diminution of the college engenders a feeling of distance  from the University; being sold a product is very different to sharing.

REF impacts not only on direct governmental funding and student numbers --- it is also  expensive to run. 
According to the {\emph{REF Accountability Review: Costs, benefits and burden}} ``the total cost to the UK of running REF2014 is estimated to be \pounds{246}M''. 
Thus, the stakes are very high and universities invest vast amounts of time, effort and money in trying to get the best scores they can. 
Coventry University is one part of this race of course and I next give a brief account of our experience in the Mathematical Sciences UOA.

% % % % % % % % % % % % % % % % % % % % % % % % % % % % % % % % % 
\subsection{Critical mass}
\label{2.2}
% % % % % % % % % % % % % % % % % % % % % % % % % % % % % % % % % 

Coventry University submitted  outputs (peer reviewed academic papers) of nine researchers to the Applied Mathematics  UoA for RAE2001 and, although 46th out of 58 submissions across the nation,  was ranked amongst the top of the so-called ``post-92'' universities (those formed after the 1992 Act).
While quite satisfactory for a university that was a leading polytechnic, it was clear that the research directions in Coventry were ``fragmented''. 
It was against this backdrop that I was hired in 2002.

In 2005 the University put out a call to establish new research centres and this led to the formation of the Applied Mathematics Research Centre (AMRC), which many readers of this journal will be familiar with.
{{The recruitment}} of Christian von Ferber in 2007 
marked the establishment of the Statistical Physics Group
{{and}}
with a strong Fluid Dynamics sister group, and polymath Robert Low (whose knowledge of mathematics and physics is almost universal) to bridge the gap, seven stalwart applied mathematicians optimistically marched  into RAE2008. 

\begin{figure}[!t]
\centerline{\includegraphics[angle=0, width=5.9cm]{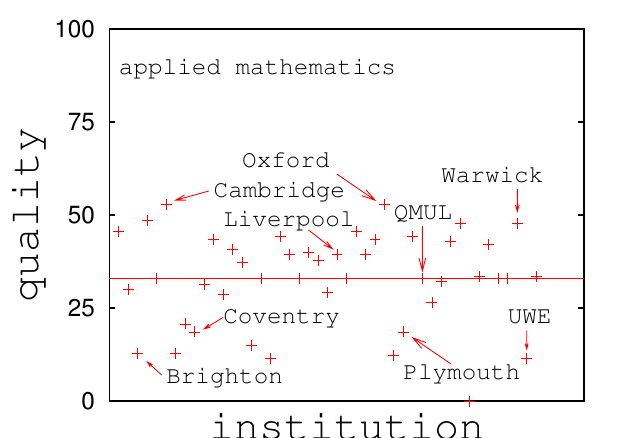}
          \includegraphics[angle=0, width=6.0cm]{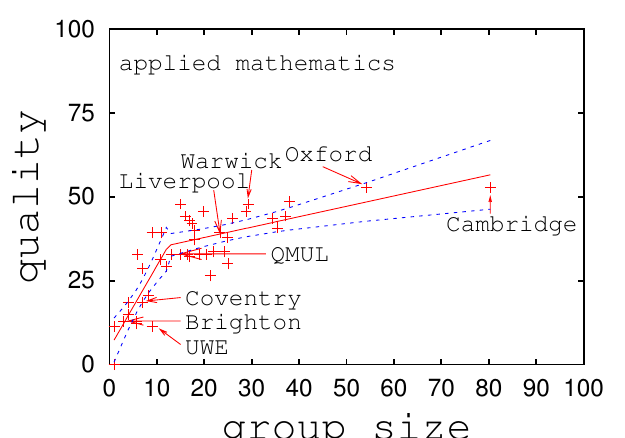}}
 \caption{(Colour online) 
RAE2008 results for 45 applied mathematics groups.
Left: quality measures plotted against alphabetically listed university names.
Right: quality plotted against sizes of submitted research groups. 
The ``phase-transition'' evident led to quantification of the notion of critical mass and suggested support for pockets of excellence as an optimal national strategy.
}
\label{fig3}
\end{figure}

The outcome of that exercise is summarised in  figure~\ref{fig3}. 
The vertical axis of the left panel denotes a measure of quality of research in applied mathematics and the horizontal axis is simply a list of 45 UK universities ordered alphabetically.
The reader is referred to the original literature for details \cite{1=44=EPL}.

Coventry was ranked below average, represented by the horizontal red line on the left panel of figure~\ref{fig3}. 
But this position  masked the accomplishment of the new group,  for, of over 
100 universities nationwide, only 45 were able to submit in Applied Mathematics at all. 
The AMRC did remarkably well given the short time it had to establish trust and cohesion after a very competitive and bitter internal struggle to get off the ground 
(one is reminded of the Wall Street Journal's famous 1973  quote of Sayre's law: ``Academic politics is the most vicious and bitter form of politics, because the stakes are so low'').

However, viewed with the eye of a statistical physicist, quite a different story is hidden behind the data.
The {{vertical}} axis of the right panel of figure~\ref{fig3} is the same as in the left panel and the {{abscissa}} is the size of the submission --- e.g., Coventry submitted 7 staff in Applied Mathematics in comparison to Oxford's~54. 

Similar plots for other UOAs show similar patterns and the reader is again referred to the literature~\cite{2=45=KeBe11}.
The breakpoint  in figure~\ref{fig3} is akin to a phase transition and invites analysis.
This was the subject of extensive research with Bertrand Berche in Lorraine in the period following RAE2008~\cite{1=44=EPL,2=45=KeBe11,3=46=Normalization,4=51=OECD,5=52=manage}.
We considered research groups as complete networks of interacting people and developed a mean-field model that takes interactions into account.
The concept adds statistical physics to Robin Dunbar's observations of  patterns in human group sizes and the reader is referred to \cite{Dunbar} for a fascinating account.

\looseness=-1 Bertrand and I published critical masses of over 25 academic disciplines.
This is  approximately half of the breakpoint position in plots like figure~\ref{fig3} and is the smallest size needed for a group to be viable in the long term. 
This research impacted on national and international professional bodies such as 
{{statistical~\cite{6=53=statisticians}, educational~\cite{A2=BERA}, mathematical~\cite{A3=IMA} and  physical societies \cite{A4=Reflet}.}}
Further impact was on {\emph{Times Higher Education}} and on organisational structures from The Foundation for Science and Technology to Verisk Analytics. 
But it also impacted onto the statistical physics group and on Coventry itself. 
Accepting evidence-based research, the University agreed that in order to shift its position in the left panel of figure~\ref{fig3} it has to move to the right in the right panel. 
This led to the hiring of new staff in the AMRC. 

Critical mass was achieved and, although only 9 staff were submitted, over 15 contributed to mathematical sciences research --- not least through the trusted links in Coventry's fully connected network.
This selectivity was a tactical move by senior managers who were more interested in rankings (to attract ``customers'') than funding that would emanate from the exercise. 
It also reflects the change from RAE to~REF. 

But of course the theory relies on the assumption that all interactions between academics in a group are positive. 
In \cite{5=52=manage} we suggested that a strong degree of cohesiveness (trust) lies behind the success of the best  groups, citing applied mathematics and physics as amongst the most prominent examples. 
Success involves optimization of the quality of communication links between research group members.
Conversely, violation of such links is a clear impediment to the success of a group.
Academic rivalry is a nasty phenomenon and plagiarism of ideas is its most ugly and despicable manifestation.
Even some of China's top scientists can be prone to scientific misconduct \cite{China}.
We have yet to tackle negative links in the network assumptions of the above literature and we will disclose its effects in a future publication.
Suffice to say at this point that iterated prisoners dilemma models show that, while academic impropriety  may be beneficial for the culprit in the short term,
it is not in the long term and payback is inevitable~\cite{Tim}.

\begin{figure}[!t]
\centerline{
\includegraphics[angle=0, width=5.8cm]{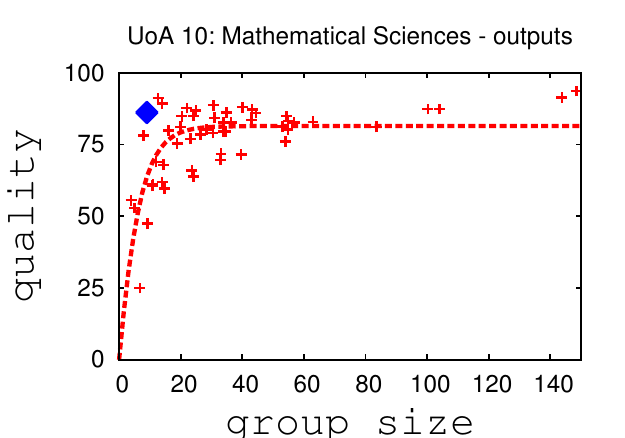}
\includegraphics[angle=0, width=5.8cm]{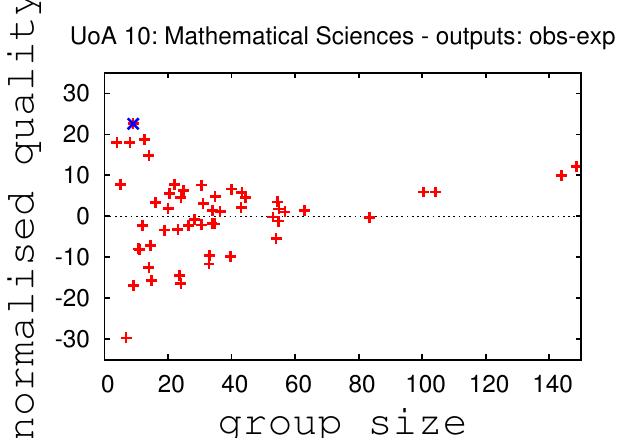}
}
 \caption{(Colour online) 
REF2014 results for 53 mathematical sciences groups.
Left: quality measures (percentage of world leading or internationally excellent papers) plotted against sizes of submitted research groups. Logistic curve indicates size-dependency of average qualities. 
Right: quality normalised against logistic curve indicating extent to which each group punched above their weight (quality minus average quality for groups of the same size).
The large blue symbols represent Coventry's mathematical sciences group.
\label{fig4}}
\end{figure}

Our next challenge was REF2014 and  figure~\ref{fig4} captures  Coventry's performance.
With 87\% of its submitted papers rated as world leading or  internationally excellent (4* and 3* in REF parlance) Coventry was ranked 12th out of 53 submissions in
Mathematical Sciences in terms of this metric.
The extent to which Coventry punched above its weight  is  clearly illustrated in the right panel where the data are normalised according to submission  size  (actual performance minus expectations of the logistic curve of the left panel).
Coventry ranks highest according to this metric --- clearly we are best at inventing metrics in which we are best!

We also did well in terms of impact --- 80\% internationally excellent (3*) scoring 24th out of 53 nationally.
Half of the impact that achieved this score for REF2014 was based on our study of RAE2008 discussed above.
We decided to do a similar ``trick'' (in the ``mathsy'' sense of the word) for  REF2021 and predicted the results of REF2014!

% % % % % % % % % % % % % % % % % % % % % % % % % % % % % % % % % 
\subsection{Predicting REF}
\label{2.3}
% % % % % % % % % % % % % % % % % % % % % % % % % % % % % % % % % 

This is where we required the expertise, enthusiasm and spirit of the institute that Ihor directs --- in particular Ihor's daughter Olesya Mryglod, then already an accomplished and independent young scientist in her own right, and Yurij Holovatch, Head of Department for Statistical Theory of Condensed Systems. 
At the time (around 2014) there were ``powerful currents whipping up the metric tide [including] 
demands by policymakers for more strategic intelligence on research quality and impact; 
competition within and between institutions for prestige, students, staff and resources
\ldots
and the capacity of tools for analysing them''~\cite{MT}.
Replacing peer review in future REFs by metrics would have  reduced the cost of the expensive and time-consuming exercise
and place it firmly in the hands of professional services with no expertise required to press a button and decide the future of entire cohorts of academics (and their families, friends, etc.).

\begin{figure}[!t]
\centerline{\includegraphics[angle=0, width=5.5cm]{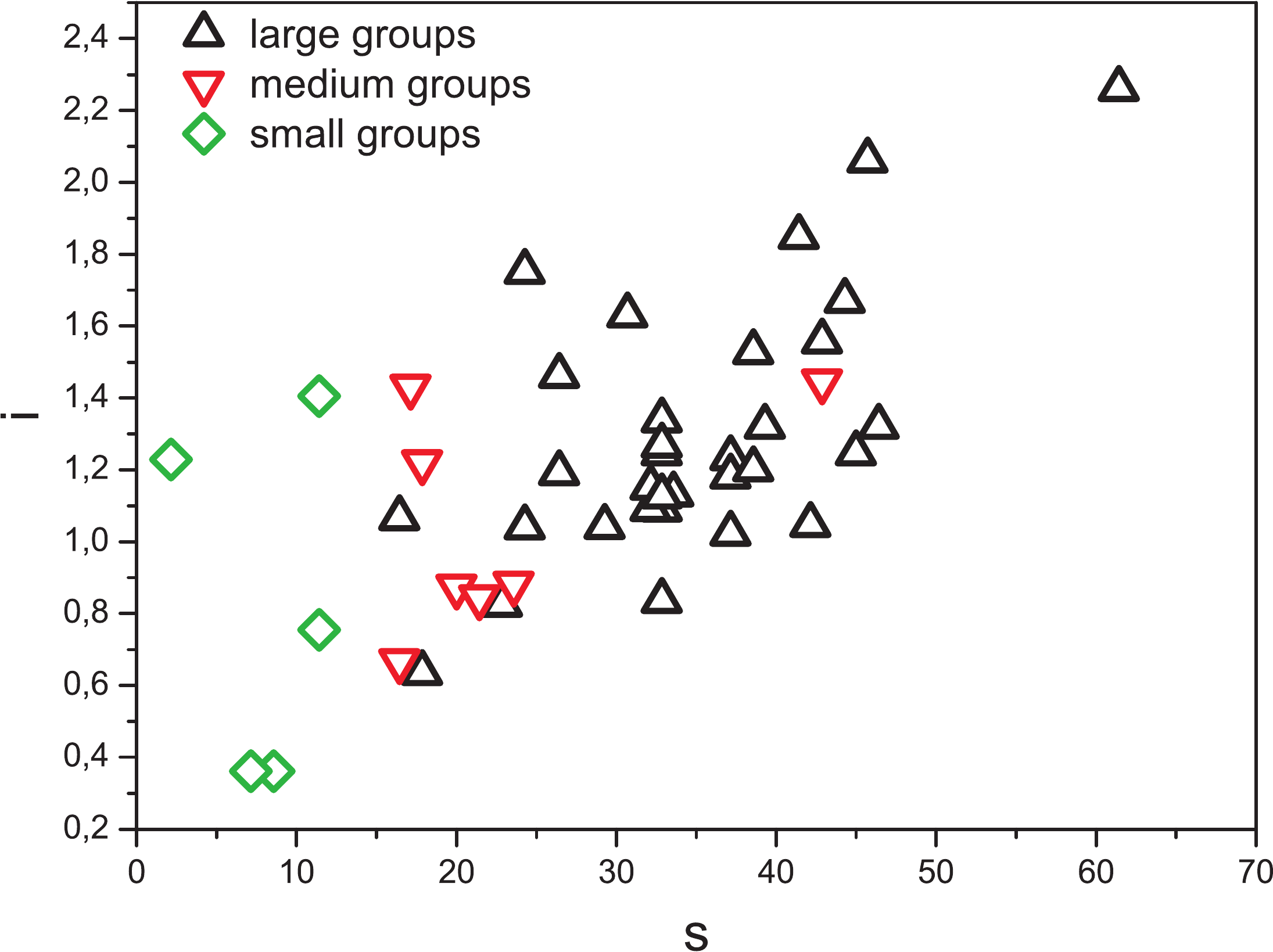}}
 \caption{(Colour online) 
Correlation between peer review measures of quality and scientometrics indicators for biology research groups.
The vertical access, labelled $i$, draws from a metric known as the normalised citation impact and the horizontal axis, labelled $s$, comes from RAE2008 peer-review results. 
For definitions of $i$ and $s$, the reader is referred to \cite{7=54} and critical-mass definitions of ``small'', ``medium'' and ``large'' 
are  from \cite{2=45=KeBe11,4=51=OECD}.}
\label{fig5}
\end{figure}

To investigate the feasibility of this on a grand scale, we decided to predict the outcome of REF2014 using citation counts. 
We first performed a retrospective analysis of REF2008 by examining how its outcomes correlated with citation-based indicators~\cite{7=54,8=58}.
We found good correlations for overall or aggregate measures of group strength but poor correlations for per-head measures of  quality (see figure~\ref{fig5} for biology). 
This is explained by the difference between intensive and extensive measures (to use language of thermodynamics) and 
means that while citation counts might possibly be used to decide funding, they should not be used for ranking  quality \cite{7=54}.
However, the situation is even more nuanced than that --- we also found that any predictive correlations that do exist are for large teams only \cite{8=58} (size being defined in \cite{1=44=EPL,2=45=KeBe11,3=46=Normalization,4=51=OECD,5=52=manage,6=53=statisticians,A2=BERA,A3=IMA,A4=Reflet}).

Having established that size matters in research, we went on to ``predict'' the rankings for REF2014 
and placed our predictions in the public domain before REF's results were published~\cite{10=80}.
Our  paper ``drew considerable interest in the media and on the blogosphere'' and we revisited our forecast after the official REF results were released~\cite{11=81}.
``Our predictions failed to anticipate with any accuracy either overall REF outcomes or movements of individual institutions in the rankings relative to their positions in the prevous RAE (2008).''
We concluded that ``we feel an extended discussion on the role of metrics in national research assessment exercises of the types considered here is warranted''~\cite{11=81}.

%%%%%%%%%%%%%%%%%%%%%%%%%%%%%%%%%%%%%%%%%%%%%%%%%%%%%%%%%%%%%%%%%%%%%%%%%%
\subsection{What expertise can we trust? Calibrate with confidence}
\label{2.4}
%%%%%%%%%%%%%%%%%%%%%%%%%%%%%%%%%%%%%%%%%%%%%%%%%%%%%%%%%%%%%%%%%%%%%%%%%%

\begin{figure}[!t]
\centerline{\includegraphics[angle=0, width=6.5cm]{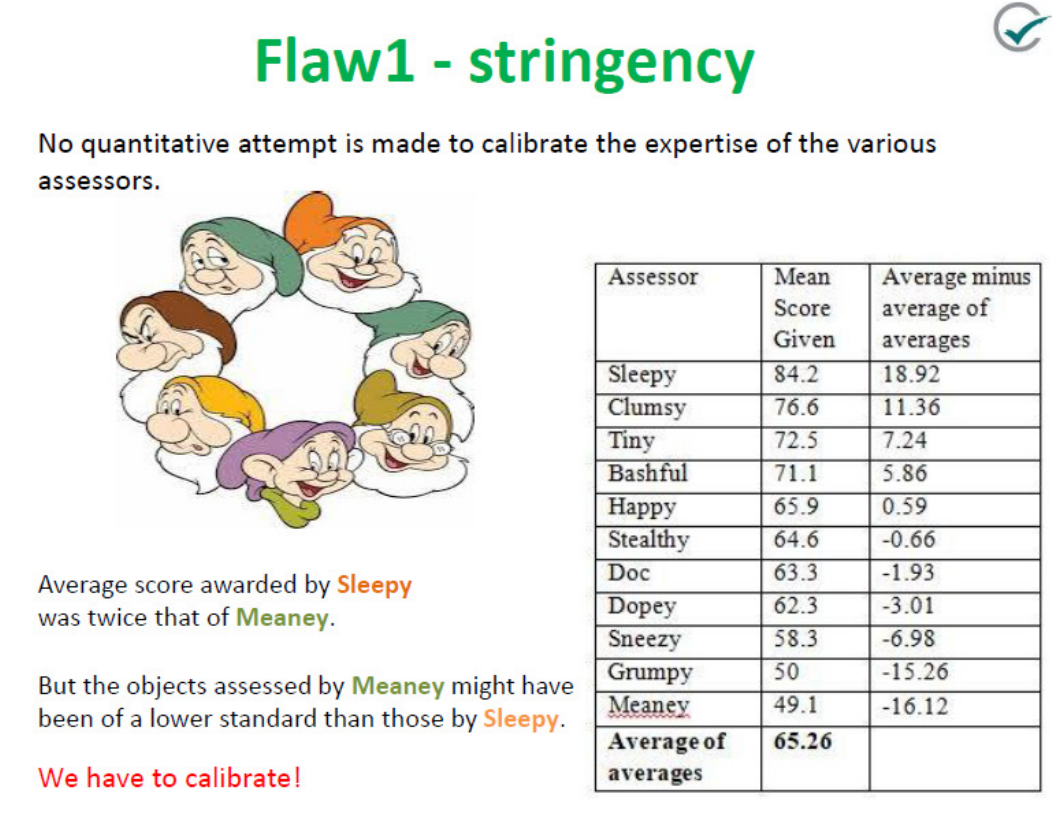}
%            ~~~~~~
            \includegraphics[angle=0, width=6.5cm]{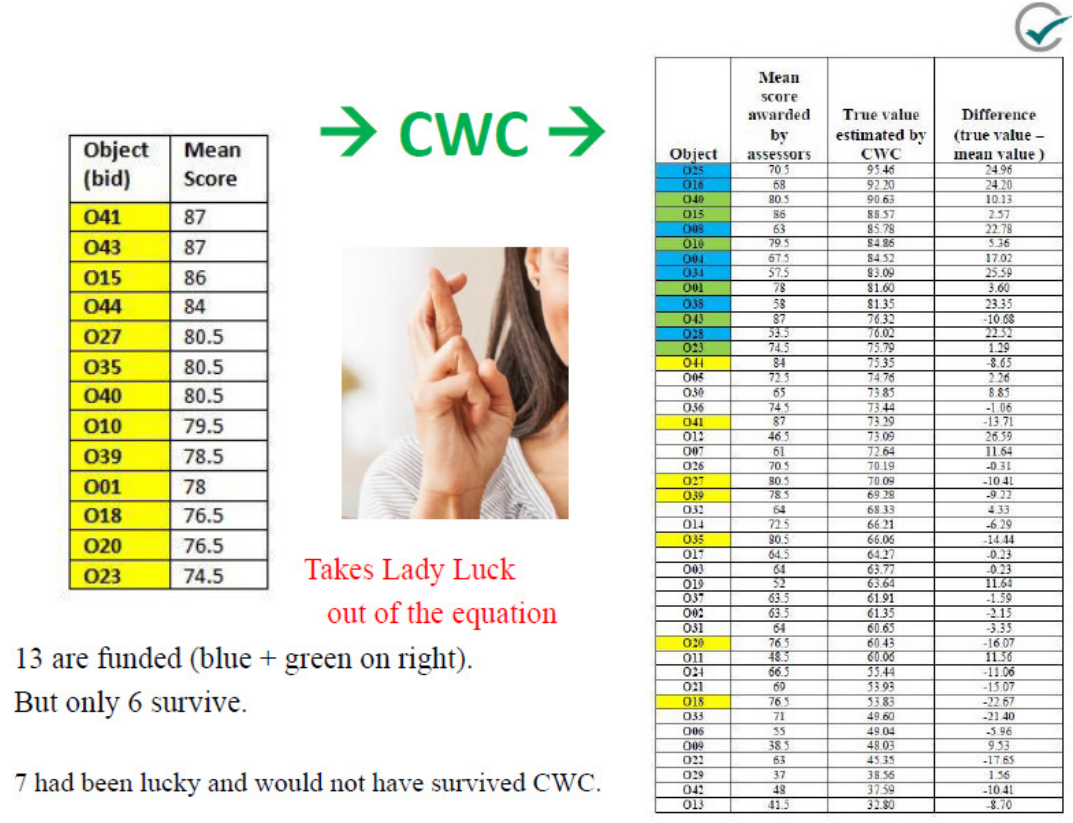}}
 \caption{(Colour online) 
Presentation slides for Calibrate with Confidence. 
Left panel: average scores given by anonymized  assessors. 
Right panel: Dramatic effect of the algorithm. 
For details of the experiment refer to \cite{17=101=CWC}.}
\label{fig6}
\end{figure}

{Our} work  drew the attention of noted academics. 
Robert MacKay had drawn from his experience as an RAE panel member, where he was struck by the need for calibration of assessors. 
He developed an algorithm to estimate assessor-stringencies and ``true'' scores from raw scores and uncertainties. 
We joined forces and tested the method on  real and  simulated data.
Some data came from Coventry University's competition for internal funding which involved 44 bids and  11 assessors. 
As in assessment processes worldwide, it was impractical for each assessor to assess each bid.
Instead, 
each bid was assessed by more than one assessor and averages were taken. 
The panel convened, discussed, and ranked the bids.
The resulting data were anonymised in \cite{17=101=CWC} and are presented in a more user-friendly form in figure~\ref{fig6}.
Each panel is an image from presentations I have given internally and externally. 
The right side of the left panel gives the average scores given by the 11 assessors.
Some  appear more parsimonious than others --- the average mark given by ``Meaney'' is about half that of ``Sleepy'' 
(I can only reveal one assessor's identity; I myself am ``Bashful'').
But we can't simply shift all marks by a given amount to match their averages; 
it could have been that the bids assessed by Meaney were  genuinely poorer in quality and that Sleepy was given good proposals to assess.
Clearly proper calibration is needed. 

We again refer the reader to \cite{17=101=CWC} for mathematical details  but, suffice to say, Calibrate with Confidence takes Lady Luck out of the equation (see also \cite{Rob}).
The yellow items on the right panel are the 13 of the 44 bids that were ranked highest and funded under the standard system.
The blue ones are those bids that failed under the standard scheme but would have been funded had the algorithm been used. 
The green bids would have succeeded under either scheme.

For example, the bid identified as O18 was successful under the standard scheme but was assessed by assessors labeled Clumsy and Tiny (listed in figure~\ref{fig6}) who were over-generous.
At the opposite end of the scale O25 was unsuccessful, having been assessed by miserly  Grumpy and Sneezy.
Confidences also played a role; O39 was successful but had high mark from Sleepy in whom confidence  was low.

This process clearly demonstrates the role luck plays in standard processes used worldwide. 
Calibrate with Confidence tackles this glaring flaw in panel assessments and its potential for impact inside and outside  academia is obvious.

% % % % % % % % % % % % % % % % % % % % % % % % % % % % % % % % % 
\subsection{Summary of Coventry's scientometrics work}
\label{2.5}
% % % % % % % % % % % % % % % % % % % % % % % % % % % % % % % % % 

In summary, with trusted partners, Coventry's statistical physics group extended its activities into scientometrics --- especially (a) critical mass, (b) correlations (metrics vs. peer review) and (c) Calibration with Confidence. 
With friends from the fluid dynamics group (19 researchers in total), we  march confidently into REF2021.
(In the meantime, more University restructuring led to the establishment of the Fluid and Complex Systems research centre which adsorbed the AMRC.)
However, you will recall that impact is now an essential element of the journey --- it counts as 25\% of the overall score (with outputs and environment weighing in at 60\% and 15\%, respectively).

%%%%%%%%%%%%%%%%%%%%%%%%%%%%%%%%%%%%%%%%%%%%%%%%%%%%%%%%%%%%%%%%%%%%%%%%%%
\section{Impact of trust: Collaborating with Coventry}
\label{3}
%%%%%%%%%%%%%%%%%%%%%%%%%%%%%%%%%%%%%%%%%%%%%%%%%%%%%%%%%%%%%%%%%%%%%%%%%%

REF2021 demands submission of {\emph{impact case studies}} which describe non-academic impact of excellent research.
With under 20 researchers in mathematical sciences, Coventry requires two such case studies.
One of these involves the multidimensional, rich, and joyful impact that spontaneously arose out of curiosity-driven mathematical forays into comparative mythology --- a project we refer to as ``{\emph{Maths Meets Myths}}'' (MMM).
Since my submission to this Festschrift is to honour Ihor's contribution to research structures rather than to blow the trumpet for MMM, I defer that story to another day.
Instead I focus here  on  our second impact case study which is underpinned by work with Lviv and pertains to research structures.
But first I describe  impact with  Lorraine.

% % % % % % % % % % % % % % % % % % % % % % % % % % % % % % % % % % % % % % % % 
\subsection{Impact from Coventry's trusted partnership with Lorraine}
\label{3.1} 
% % % % % % % % % % % % % % % % % % % % % % % % % % % % % % % % % % % % % % % % 

At the time of our first forays into the world of scientometrics \cite{1=44=EPL}, UK policy was that funding should be focused on bigger universities with bigger groups.  
In 2011 {\emph{Times Higher Education}} reported 
``The research councils \ldots have confirmed that more research funding will be concentrated on top-performing universities, with resources \ldots selected on the basis of size \ldots''~\cite{Colquhoun}.
The eminent  and influential pharmacologist  David Colquhoun, for example, 
was vocal in his opinion that even with this policy there was ``not enough concentration for UK research to fulfill its potential''. 
He advocated ``confining all research activity to pre-1992 universities'' (i.e., excluding the likes of Coventry) and ``to convert a proportion of \ldots post-1992s  \ldots  into teaching-only institutions''~\cite{Colquhoun}.
Our  critical-mass model  (described in section~\ref{2.2} and documented in \cite{1=44=EPL,2=45=KeBe11,3=46=Normalization,4=51=OECD,5=52=manage,6=53=statisticians,A2=BERA,A3=IMA,A4=Reflet}) was in direct contradiction to this  policy and it triggered a robust debate in media and elsewhere.

So who won?

In 2018, Research England, allocated {\pounds}75,000,000 to support ``pockets of excellence''~\cite{E3} using precisely the term used in at least three of our papers \cite{1=44=EPL, 2=45=KeBe11, 5=52=manage}.
{But is this a coincidence or was there impact?}
Our work had {certainly} impacted on lobbying organisations, professional associations and the media to inform and change policy on research quality~\cite{usREF2014}.  
For example, \emph{Research Professional} wrote that our work ``found that after a point, increasing the number of researchers did not add proportionately to quality of the work, suggesting medium-sized research groups are preferable to concentrating research in a smaller number of large groups''~\cite{RP}. 
However, this easily-evidenced flurry of media activity {was a long time ago.}
{People} move from one role to another and it becomes harder to pin down the precise chain of any influence that occurred. 
Ironically, the more accepted {ideas} become, {even evidence based ones,} 
the harder it is to {prove} causality retrospectively.
Can we claim this impact for REF2021?

Calibration notwithstanding, a positive person like Sleepy or Happy might say ``yes'' (I know who they are and they have confirmed they would do so). 
A less trusting, negative person would more likely say ``no'' (I know Meaney and Dopey too). 
Clearly, we cannot rely on Lady Luck and need better evidence of impact.

Maintaining and gathering evidence for impact arising from older work is extremely time consuming.
One might think it should fall within the remit of professional services.
Leaving it to academics impedes further research and that can scarcely be the intention of REF.
However, anyone performing this labour intensive work needs to be intimately familiar with the research and that requires engagement right from the outset.

% % % % % % % % % % % % % % % % % % % % % % % % % % % % % % % % % % % % 
\subsection{Impact from Coventry's trusted partnership with Lviv}
\label{3.2} 
% % % % % % % % % % % % % % % % % % % % % % % % % % % % % % % % % % % % 

To my delight, in early July 2015 I received an unsolicited email from Professor James Wilsdon, 
sharing with me an advance copy
of 
The Independent Review of the Role of Metrics in Research Assessment and Management 
(famously known as the {\emph{Metrics Tide}} report)~\cite{MT}.
The {\emph{Metrics Tide}}  was commissioned by Government in April 2014 and, over 15 months, investigated  roles that quantitative indicators can play in the assessment and management of research. 
James wrote
``We're very grateful for your engagement and input to this process, which \emph{enriched
our deliberations} and \emph{provoked us to ask questions} that might otherwise remain unaddressed''~\cite{Wilsdon} (my emphasis).

\emph{Enrichment} is impact; 
the Impact section of the REF Assessment Criteria~\cite{REFrules} 
states ``significance will be understood as the degree to which the impact has {\emph{enriched}} policies, understanding, awareness'', etc.
Provoking questions is also impact;
the same section gives examples of research ``which leads to an approach being {\emph{questioned}}''.
Ours is the only (``REFable'') academic work cited in the section on ``Quantitative data and the assessment of research outputs''. 
It is used there to counter arguments \cite{Dorothy}
``in favour of using the departmental h-index''
by quoting our conclusion ``that the relationship is not strong enough to justify the use of the departmental h-index as a replacement for peer review''.

Quoting a quote of oneself, as I have now twice done here, is unusual academic behavior to say the least.
{Ostentatiousness is second only to plagiarism as reprehensible in the academic world.}
But {bombasticism} is a manifestation of the bizarre borderland between the academic and the corporate.
The demand to evidence impact pits corporate crowing directly against academic abasement.
One of the ``headline findings''  of Metrics Tide  is that ``for the impact component of the REF, it is not currently feasible to use quantitative indicators in place of narrative impact case studies'' in case 
``concept of impact might narrow''~\cite{MT}.
Given that persistent adoration of quantitative indicators {in certain quarters} rides roughshod over headline findings, and given that this paper is not an impact case study, I  find a new use for {them} --- to spare me  the embarrassment of quoting more quotes of our work in Metrics Tide and associated documents \cite{MTS1,MTS2}.
Instead I invent further metrics in which we are best and I hide them in 
archives available on request.
Suffice to say here that our {papers are the most cited, most quoted and and the most closely aligned to Metrics Tide itself. The final sentence in the final page of the main body of the report announces a ``Bad Metric'' prize to be awarded to ``the most egregious example of an inappropriate use of quantitative indicators''.  I believe the prize has not yet been awarded but by metricising our impact on Metrics Tide I might be in with a chance! Especially as our works} and Metrics Tide both conclude that metrics cannot replace the peer-review process in research assessment and that it is not currently feasible to assess research outputs or impacts in the REF using quantitative indicators alone. 
{But this time, unlike the previous section, we have more than correlation --- unsolicited, repeated, quantifiable evidence of direct causation.}
As the work with Lorraine earlier, our work with Lviv (Mryglod et al.) had initially met resistance from the ``powerful currents whipping up the metric tide''.
Therefore, apart from the delight of receiving this degree of attention from esteemed academics, it was very rewarding to have our points vindicated in {and impacting on} an enormously powerful report of national and international importance.

{{Verification of impact {can be} very hard to gather --- even impact emanating from Metrics Tide itself!}} 
{{For}} example, at recent presentations in Coventry, a senior customer consultant at Scival (an Elsevier company that ``offers comprehensive access to the research performance of over 14,000 research institutions{'')}
cited it {{multiple}} times to demonstrate awareness of ``responsible use of metrics''~\cite{Walker}.
However, my request for a statement that Metric Tide impacted on Scival was not welcomed in case I misquoted it 
--- an example of the distrust between ``them and us''.
A complete gathering of all  indirect impact is impossible  given its vastness (and difficulty evidencing it) and we can only gather a few sample testimonials 
{for} a REF submission.

But the impact {{of Metrics Tide}} is far more significant and wider reaching than this. 
{\emph{Building on Success and Learning from Experience}} is an  independent review of the REF (commonly referred to as the {\emph{Stern Report}}), {comm}issioned by the 
{UK government in} 2015 and published the following year~\cite{Stern}. 
It forms the basis for much of REF itself.  
Lord Nicholas Stern accepted the 
Metric Tide
{and} recommended that peer review remain the primary method of assessment.
So, {{academic}} work {{that}} impacted onto  Metrics Tide
{{impacted}} 
onto the Stern Report  which helped shape the rules of REF2021 itself!
This, in turn, impacts on every university and academic in the UK and globally as  findings of Metrics Tide and the Stern Report are taken up in research assessment exercises around the world.

{One example is in Ukraine's reform of its national scientific policy.}
The Lviv System of Research~\cite{LvivSystem} is the governing council of Lviv Scientific and running a large scale  pilot  overhaul of Ukrainian scientific structures. 
It is complicated and complex and the Government recognises the importance of getting it right in order to scale it up to the national level. 
The committee became aware of our scientometrics work through media coverage and through a report for the National Academy of Sciences of Ukraine~\cite{AX=comparison}.
This is the highest research body in Ukraine as a self-governing state-funded organization.
Olesya Mryglod was invited as a key-note speaker to the workshop ``Scientometrics and research assessment: Phenomenologial, technical and practical dimensions''~\cite{Otalk} and through this the Lviv System  became aware of the Metrics Tide report and delved deeper into our research.
This led to further awareness of our earlier papers and changed their understanding of the meaning of the term critical mass in research groups.
In fact, this research findings impacted so heavily that they invited Olesya  to help formulate the shape of the Ukrainian process for the years to come.

 Over the past 4 years the Ukrainian system has been significantly impacted upon by the research of the scientometrics collaboration of Coventry, Lorraine and Lviv and they have directly confirmed that our papers have ``enriched, influenced, informed''  and indeed changed the understanding and awareness of the working group which is in the core of Lviv System of Researchers. 
Thus, the Ukrainian counterpart of REF uses the precise words of REF to indisputably demonstrate real, tangible and evidential impact that even Meany or Grumpy could not refute.

% % % % % % % % % % % % % % % % % % % % % % % % % % % % % % % % % % % % 
\subsection{Other dimensions of impact} 
\label{3.3} 
% % % % % % % % % % % % % % % % % % % % % % % % % % % % % % % % % % % % 

In this paper I primarily discuss some of the impact coming from our work with Ihor's institute and my story is in part aimed at recommendations for Ukraine's nascent counterpart of REF.
{An impact agenda} may well emerge as an important part of that exercise.
As I have demonstrated, evidence of impact is easier to gather as the impact occurs. 
It is far more difficult to capture as time elapses, so it is crucial to grasp it from the outset.

I am reminded of an invitation
to the Higher Education Policy Institute conference in the Royal Society London in 2015. 
%I clearly remember my introduction to a HEFCE director. 
I clearly remember my introduction to a HEFCE official.
His comment along the lines of “it is very important that you work on
this” was another boost to my interdisciplinary sojourn and my trip home from London that day was aglow with confidence that  mathematical scientists (usually so distant from society) are making a positive contribution to the world. 
My job was done and as I returned to Coventry I returned to safe world of statistical mechanics. 
Little did I realise the importance of such interdisciplinary adventures for without them we would have no impact, no REF2021 submission and hence no statistical physics in Coventry in the future.

Like perpetual motion, continuous input is required for impact to be remembered --- and hence evidenced.
And one organisation I am very pleased to continuously interact with is CAMRA --- the Campaign for Real Ale.
This is a  consumer organisation with over 192,000 members, me included,
{and a} founding member of the European Beer Consumers Union. 
Given the importance of pubs and beer for bringing different cultures together~\cite{Dunbar2}, 
we thought it an entirely appropriate application of the Calibrate with Confidence algorithm.
\begin{figure}[!t]
\centerline{\includegraphics[angle=0, width=2.5cm]{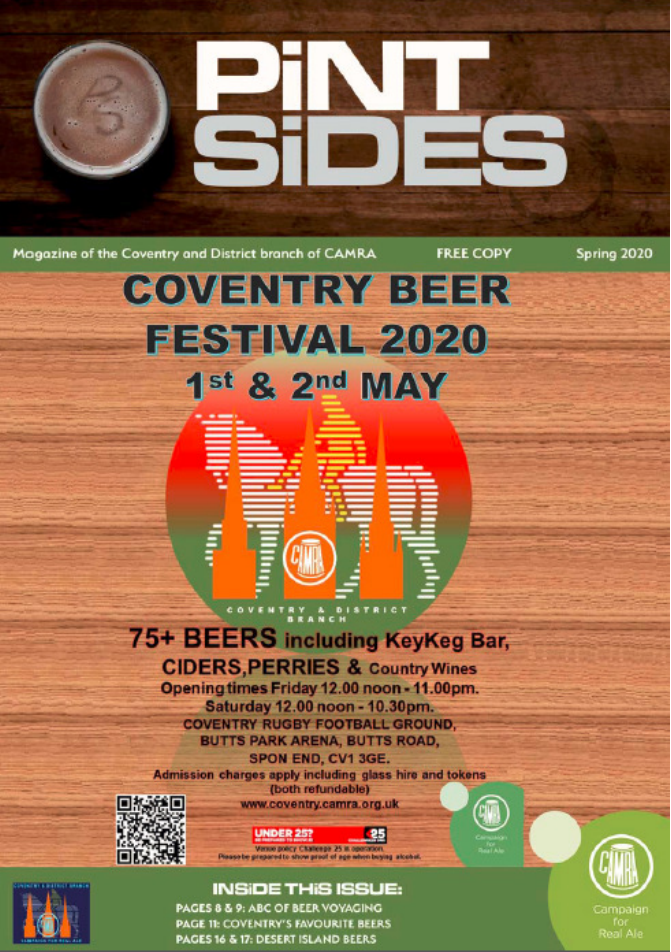}}
 \caption{(Colour online) Calibrate with Confidence {counts what counts in Coventry.}
}
\label{fig7}
\end{figure}
We have 
already trialed Calibrate with Confidence on CAMRA data {{(see figure~\ref{fig7}).}} 
{Their existing} Pub-of-the-Year selection processes require  every shortlisted pub to be assessed by nearly every assessor. 
The algorithm can achieve similar results without this limitation. 
Far from reducing the number of pub visits, this means more assessors and more pubs can be involved in the competition and new assessors can find out how their judgments compare to CAMRA experts~\cite{Beer}.

We have impacted on other organisations such as the Office of Qualifications and Examinations Regulation, a government department that regulates qualifications and exams.
We are now very aware that evidence of impact has to be gathered as the impact is being generated.
Although Lviv is not directly involved in these two impact-generation projects, we are quite willing to  calibrate  pubs in Ukraine and elsewhere. 
(We are nearly as happy to calibrate exams but also
happy to refer interested parties to our paper or website~\cite{17=101=CWC}.)
After all, without the input of Ihor's team much of this research would not have been carried out so the indirect impact is also theirs.

%%%%%%%%%%%%%%%%%%%%%%%%%%%%%%%%%%%%%%%%%%%%%%%%%%%%%%%%%%%%%%%%%%%%
\section{When is impact impact? When cultures clash?}
\label{4} 
%%%%%%%%%%%%%%%%%%%%%%%%%%%%%%%%%%%%%%%%%%%%%%%%%%%%%%%%%%%%%%%%%%%%

It seems obvious to me that the above impact is rich in reach and significance for REF2021.
Therefore, for the third time (after RAE2008 and REF2014), I look forward to submitting to REF so that Coventry and collaborators can continue producing world leading and  internationally excellent research for years to come.
A successful submission will convince that {we can be trusted} to get on with things. 

However, it is not as simple as that --- to have a  REF submission at all we, {{as academics in other universities,}} have to get past the vetting process of professional services to convince that our impact is world leading or  internationally excellent!
But what if the internal process is mistakenly stringent or overly linear?
We have to get the camel through the eye of the professional-services needle to get to the kingdom of academic peer review and the world of academic freedom that undoubtedly lies beyond.
What if 
authorities view impact with the wrong eye?
Professional services are not calibrated  like the trained independent thinkers on the REF panel.
Indeed, independent thinking of the type nurtured by academic freedom is not exactly encouraged in ``top-down'' sectors of the corporate world. 
Like spins in a cool ferromagnet seeking to minimise free energy, professional services prefer alignment for equilibrium~\cite{Bankers}.
And academics themselves are cool to the idea of being distracted  
from their research, so are reliant on a small team of aligned professional services to read the rules and get everything right.
However, ``group think'' or ferromagnetic interactions can be disastrous; if 
those in charge get it wrong, wrongness may  permeate throughout an entire 
system \cite{CaC}.
They may mistakenly deem we are ``not meeting REF targets' and, therefore, have to be closed!''~\cite{Covtel}.

% % % % % % % % % % % % % % % % % % % % % % % % % % % % % % % % % % % % 
\subsection{Endless debates --- what is impact?}
\label{4.1} 
% % % % % % % % % % % % % % % % % % % % % % % % % % % % % % % % % % % % 

Here are 10 examples of statements 
I have encountered.
I give them as examples not only for those going into REF but for Ukrainian and international colleagues developing their own research assessment protocols to caution against some of the challenges we have encountered.
{Note that all impact for REF has to be ``underpinned by excellent research conducted in the submitted unit.'' The ten
fallacious examples are:}
\begin{itemize}
  \setlength\itemsep{0em} 
  \item ``Impact has to involve change'';
	\item ``Impact has to be quantifiable'';
	\item ``Public engagement  is not impact'';
	\item ``Media impact  is not impact'' (``academics are not paid to engage with the media'')~\cite{CUweb};
  \item ``Impact case studies are only measurable by non-academics'';
	\item ``Indirect impact does not count'';
	\item ``An impact case study based on a single output does not count'';
  \item ``An impact case study {{needs to be}} based on {{a}} funded research {{project}}'';
	\item ``Impact does not count if it is too parochial'';
	\item ``A paper published in a first quartile journal is world leading (4*)''.
		\end{itemize}
I know of at least one potentially {world leading} impact case study blocked on the basis of one or more of the above statements or very similar statements.
If the spins are aligned to beliefs like these, it becomes difficult to realign them --- a phenomenon akin to hysteresis.
Any attempt by a lone academic to challenge perceptions of multiple aligned authoritative parties
%set to right 
can be met with hysterical reactions; anti-ferromagnetic interactions are not welcome in a ferromagnetic world.
Contrast the above with the actual rules of REF~\cite{REFrules}:
\begin{itemize}
\setlength\itemsep{0em}
  \item Paragraph 192: 
``Significance will be understood as the degree to which the impact has enabled, {\emph{enriched}}, {\emph{influenced}}, {\emph{informed or}} changed the performance, policies, practices, products, services, {\emph{understanding}}, {\emph{awareness}} or well-being of the beneficiaries'';
	\item Paragraph 309: ``Impact case studies will refer to a wide range of types of evidence, including {\emph{qualitative}}, quantitative and tangible or material evidence, as appropriate'';
			\item Paragraph 286: ``Public engagement may be an important feature of many case studies, as the mechanism by which the impact claimed has been achieved'' (see also \cite{Northumbria});
	\item Annex A: Impact indicators include ``citation by journalists, broadcasters or social media'';
	\item {The bulk of}  assessors for Mathematics {are academics and all} had doctorates in REF2014;
		\item Paragraph 286: ``The relationship between research and impact can be {\emph{indirect}} or {\emph{non-linear}}'';  
		\item Paragraph 319: ``Each case study must \ldots include references to \emph{one} or more key research outputs'';
	\item Paragraph 329: Any funding data ``will enable research funders to track and
evaluate the impact of their funding. It will {\emph{not}} be provided to the panels and will {\emph{not}} form part of the five-page limit for impact case studies'';
		\item Paragraph 297: ``Impact \ldots may be {\emph{local}}, {\emph{regional}}, national or international'';
  \item Paragraph 250: ``{\emph{No sub-panel}} will use journal impact factors \ldots in their assessment''.
		\end{itemize}

Besides blatant wrongnesses like some of the above I have heard incredible comments  --- comments laden with distrust of the obvious --- skepticism  and cynicism taken to an extreme. 
For example, a senior {professional}
(who has now left the University to profess their professionalism elsewhere)  once asked how we can assume that the million people whose TVs were tuned in when I was talking about {maths and myths} (MMM) weren't just eating their dinners!
The presumption that one million people might have turned off precisely at the moment I was  on air  is one I hadn't thought of. 
It is somewhat disconcerting but, it must be admitted, not impossible.
I could take solace from face-to-face comments such as when, for example, the Princess Royal came to town and I was asked to present on the same topic.
Comments like  ``What a charming Irishman, I hadn't realised one could do such interesting things with maths''  might 
{{or might not}}
boost {{one's}} ego   but are hard to capture in writing.
I didn't think it diplomatic to present a ``Who, What, Where, When, Why, How'' question sheet that professional services use to ``measure'' changes in understanding and awareness.
The extreme version of cynicism  flies in the face of academic rationalism or, indeed, common sense. 
Paragraph 306 of \cite{REFrules} states ``attendance figures \ldots may \ldots provide an indication of the reach of the impact [and] significance \ldots might be demonstrated, for example, through participant feedback or critical reviews''.

I would imagine that {\emph{Physics World}} commissioning us to help write an article on our research is convincing as impact. 
The fact that the editors deemed it worthy of {\emph{six}} pages is even better!
Better still is that it featured as the main item on the {\emph{front page}}.
Therein, to my mind at least, lies the significance. 
The reach is indicated by the fact that this monthly magazine of the Institute of Physics has 110,000 readers of diverse backgrounds including ``industry, academia, the classroom, technician roles or in training programmes as an apprentice or a student'' \cite{IOP}. The Institute of Physics itself declares ``our {\emph{reach}} goes well beyond our membership to all who have an interest in physics and the contribution it makes to our culture, our society and the economy'' \cite{IOP}.
Yet professional services insist on contacting the journalist to ask for evidence we had ``impact'' --- as if to suggest nobody read that particular issue or everyone that did already knew about our applications of maths to myths.

I would likewise imagine that coverage in {\emph{Chemistry World}} would  count as impact.
The Royal Society of Chemistry is the world's oldest chemical society and the largest in Europe. 
{\emph{Chemistry World}}, sent monthly to all members, is Europe's largest chemistry magazine.
That it has 50,000 members should indicate reach.
The fact that not one but {\emph{three}} articles on our research were published over a five month period would indicate that at least the first two were impactful and not a fluke --- it has significance! 
The fact that the science and business journalist who wrote the articles has twenty years experience in ``writing for a wide range of publications including {\emph{Chemistry World}}, {\emph{New Scientist}} and {\emph{The Times}}''  might indicate she knows what she is doing. 
Yet professional services think we should hassle her for a testimonial.

I would imagine {\emph{three}} articles in {\emph{The Times}} newspaper (circulation of 446,164) would likewise indicate that at least the first was not a fluke. We have far more media impact in the MMM case study than these examples I give. 
But I don't see that it has to be published more than three times in the same outlet for a newspaper article to count as impact --- the fact that an expert journalist or editor considers our research will excite enough interest to boost their sales indicates something.
News has to be {\emph{new}} after all!

These are examples of the endless debates we have when two cultures clash. 
Although audience figures and a few testimonials were quite adequate for REF2014, they may not be stringent enough 
to get through the eye of the needle to submit to REF2021.
If the rules of REF are 
not clear enough, 
I propose to summarise the essence of some that seem to cause most confusion.

% %  % % % %  % % % %  % % % %  % % % %  % % % %  % % % %  % % 
\subsection{Fundamentals of the laws of impact}
\label{4.2}
% %  % % % %  % % % %  % % % %  % % % %  % % % %  % % % %  % % 

As the laws of thermodynamics {cited by C.P.~Snow} appear to come from ``on high'' but require statistical physics to fully grasp them, 
so too do the rules of impact come from ``on high'' but require common sense to grasp them.
My inspiration for writing about this comes from endless debates with professional services about the nature of impact in the media in particular.
Put simply, my argument is that talking to a million people for five minutes on TV is not the same as nothing.
Back-up for my argument includes  impact case studies that were almost certainly rated world leading for mathematical sciences at REF2014 --- e.g., that of Roger Penrose \cite{Penrose} or Marcus du~Sutoy \cite{duSutoy}. 
These were based on ``writings, public lectures and media appearances'', ``TV, radio, public lectures, social media and interactive projects''. 
As in these cases, reach in my case is easily quantified
by audience figures and significance by repetition or representative testimonials  
(to prove not everyone eats their dinner when I am on TV).

The first law of thermodynamics identifies internal energy as the sum of  work done and heat added.
Work involves movement --- something has to change.
Likewise the impact cherished by professional services  involves tangible, measurable change.
From statistical physics we know that adding heat to a macroscopic system excites the microscopic particles which constitute it --- but we don't have to  quantify these changes at microscopic level. 
Likewise perceptual change or  enlightenment via media engagement excites the neurons in the brain 
but we shouldn't have to measure every single neuron in microscopic detail to be absolutely sure.
So, the first law of impact can be that impact involves benefit as well as change --- i.e., media impact is impact!

\begin{figure}[t]
\centerline{\includegraphics[angle=0, width=4.5cm]{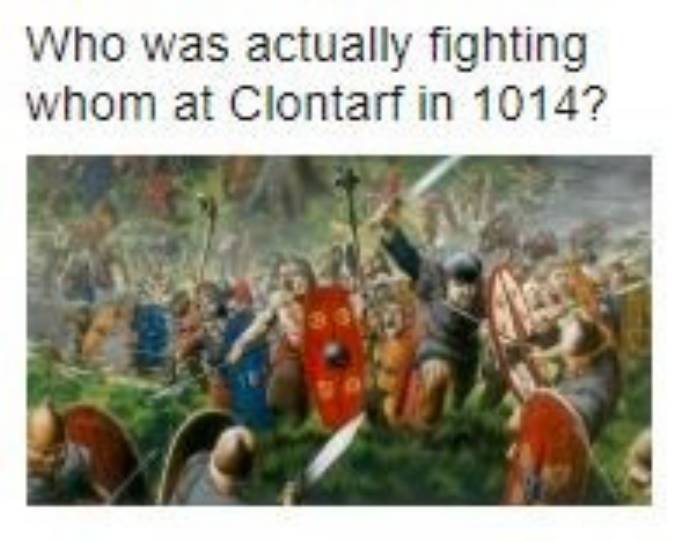}
\includegraphics[angle=0, width=4.5cm]{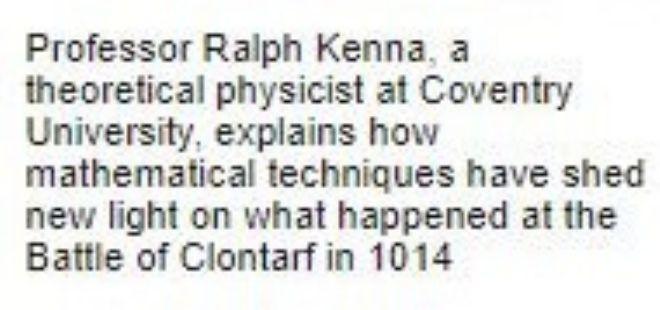}}
 \caption{(Colour online) 
Whether my interview on  national radio sucked EUR 1 million out of the Irish economy or enriched people's lives might depend on whether you are a professional or a professor.}
\label{fig8}
\end{figure}

People have tried ``come up with measurable indicators'' of media impact. In his Research Impact Handbook, Mark Reed suggests the struggling researcher could determine ``the readership of a particular newspaper or online media source and how might this translate into money via advertising
revenues linked to page views and circulation''~\cite{Reed}. 
However, he also points out that ``this approach makes many assumptions (e.g. that \ldots the coverage is positive)''.
Since different media charge differently in different countries, {{a fan of metrics could}} instead calculate the cumulative amount of time the public engaged with a media and convert this into person-years  based on simple averages. 
For example, in 2018 my interview on a mathematical analysis of the Viking Age in Ireland (part of the {{MMM}} project) lasted 5 minutes 39~seconds on an Irish national radio programme with 423,000 listeners 
{(figure~~\ref{fig8})}.
This is the equivalent of one person listening to me continuously for 24 work years.
For the corporate minded, who believe ``time is money'', 
this converts to  \euro {1,039,917} in {average} salaries.

I didn't suggest this as an  absolute quantitative measure --- merely as an order-of magnitude comparative indicator to squeeze through the eye of the money needle so beloved by {the linear minded.}
But this argument was  rejected by a  {senior professional} who argued that time spent looking at me on TV instead of at work is negative impact!
My counter-argument  that home time is more valuable than work time fell on deaf ears. 
Perhaps, like the Copenhagen interpretation of quantum mechanics, beauty is in the eye of the beholder and  the Universe only collapses after me being on TV. 
Or perhaps a simpler concept might help.
The second law of thermodynamics tells us that entropy increases --- you can't add negative entropy. 
(Actually you can, but that's another story \cite{Petro}.) 
I propose the second law of impact is enlightenment via media is not negative.

The third law is that the entropy of a  crystal is zero at absolute zero temperature.
Here, I draw on personal experience --- after multiple attempts I now believe
that evidenceable impact will close to zero if academics don't engage and leave it entirely to professional services to ``get on with it''.
The zeroth law of thermodynamics concerns the transitivity of equilibrium. 
Impact is similar --- the impact of impact is impact even if it is hard to quantify;
even that which can't be counted counts \cite{counteccounts}.

So, there we have it --- four rules of impact from a statistical physicist's perspective.
But one should not ignore history. 
Remember Henry Adams proposal to adapt the second law of thermodynamics to predict the ultimate destruction of the world?  
To avoid such an apocolyptic 
degradation of our academic world, I propose one simple law instead --- read the rules of REF!!!
And if that is too much bother, refer to figure~\ref{fig1} and  trust us, ``we are experienced scientists''! 
\vspace{11mm}

\section{When two cultures clash}
\label{5} 

In the dim and distant past, professors (sages) held leaders (kings) to account.
Before the age of writing, the spoken word eloquently framed as poetry had the power to influence authorities for it was through this that their legacy was remembered for good or for bad.
The pen was indeed as mighty as the sword.
In Ireland, for example, this diminished through the first Norman invasions and disunity left the country ill-prepared for the second --- and we know the 
consequences of that~\cite{Marisa}!

Now we are faced with another diminishment of academia.
The very objects which we ourselves invent --- simple tools like the h-index and {{eigenfactor score}} --- are deployed as the invasion unfolds. 
These endow inexpert authorities with a misplaced sense of understanding and control.
The great danger is linear applications of one-size-fits-all yardsticks ill suited to a complex research landscape.

There are fundamental differences that inexpert {corporate professionalisation} 
of academia inflame.
For ``us'' our work is about our passions, our dreams, our identity --- our legacy to humankind.
For many of us, if we became financially independent and no longer needed to be tied to a university, we would carry on with our research anyway.
Despite being employed at four different universities throughout my career, I never worked {\emph{for}} any of them!
I work {\emph{at}} a University or better {\emph{ as part}} of the university.
This arrangement has suited us both as, while our motivations may be different, our aims coincide --- to do the best we can.
For {{most}} professional services, on the other hand, work is a means to an end --- to earn a living.
They work {\emph{for}} the university and align with the rules to achieve that end. 
Financial independence might lead to a world cruise or some other means to  enjoy the things they value.
``Legacy'' is not a commonly used term in professional services.

I feel more at home with academics from far-flung corners of the world (except vile plagarisers obviously) than with professional services at home.
We use the same words for different things --- ``theoretical'' is noble for us, it is airy fairy for the ``professionals''.
``Basic'' is foundational for us, it is simple for them.
``It's academic'' means ``it's not really important'' to them and ``that's professional'' means ``it's well done'' to us.
We use different words for the same things --- they use ``customers'' we use ``students'';
they use ``business'' we use ``university''; 
they use ``line manager'' we use ``colleague''.
The  hypothesis of linguistic relativity \cite{SapirWhorf} describes the danger --- terminologies colour attitudes; we diminish and they rise. 
If we go much further the abbreviation  ``profs'' may end up meaning ``professionals''.

And therein lies the cultural chasm --- as researchers we have dreams and we take risks to achieve those dreams.
Professional services have dreams too but they are very different --- and they dare not risk their jobs to achieve them. 
The value of risky research is now recognised by funding bodies but the burden of risk-averse bureaucracy can be a massive impediment to achieving it. 
Risk averse implies it is better to do nothing than to make a mistake; rules are rules and laws are laws.
Risk averse means accepting the laws of thermodynamics and tweaking the structure of steam engines.
Risk embracement means challenging the laws and inventing a whole new sphere of knowledge --- statistical mechanics for a quantum world~\cite{anxiety}.

\begin{figure}[!t]
\centerline{\includegraphics[angle=0, width=3.5cm]{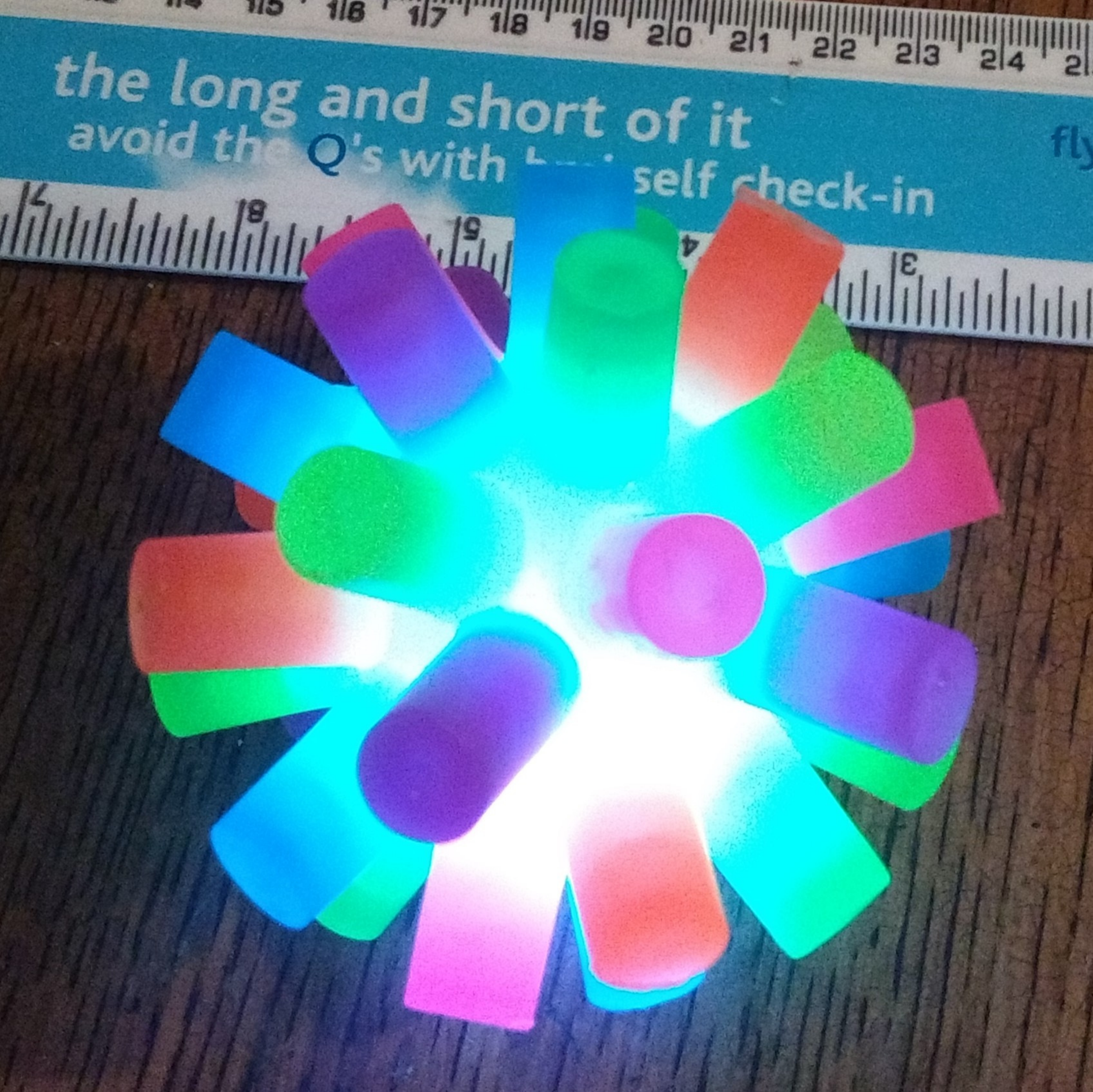}}
 \caption{(Colour online) Using a simple linear tool to measure a multidimensional vibrant case study is not the right way to count what counts --- even if it's the only yardstick you have to hand. 
}
\label{fig9}
\end{figure}

Risk averseness coupled with linear yardsticks can be a disaster --- viewed through the wrong lens, a strong multidimensional impact case study can be seen as a week linear one (figure~~\ref{fig9}).
Lord Nicholas Stern and David Sweeney recently lamented that some institutions were overly cautious in REF~2014.
They advised institutions to be bold with impact in REF 2021~\cite{SternBlog}.
As with the double-slit experiment in quantum mechanics, ``there is no simple route from one research output to one impact'' and ``the relationship between research and impact is not necessarily linear and it is not necessarily direct''.
``The very best impact submissions will likely showcase the full breadth and depth of impacts possible from the equally broad and deep range of research being undertaken in institutions, and the increased weighting for impact in the exercise will appropriately reward and incentivise further this activity.'' 

Managing multidimensional academics with multidimensional impact is a challenging task for those whose culture idolises linear, simple yardsticks.
\begin{quote}
Trying to manage anything involving academics is like trying to herd cats\ldots It means that you've got this whole group of people who are all independent thinkers and will do things if they think it will suit them \ldots but you know, they won't do it just because you say so.
\end{quote}
This is the view of one senior administrator as quoted in \cite{Herding} (another worthwhile read).

We are indeed like cats --- Schr{\"{o}}dinger's cat, confident of our vivacity inside the box. 
Like thermodynamicists they sit bewildered outside with a linear yardstick  trying to measure us with austere understandings of the impact world.
Impact has to involve ``change'' preferably quantifiable as profit;  no public engagement, no media, maximal cynicism.
Perhaps it would be better, they say, not to submit a multidimensional case study to maths but instead take the slice of our impact that is measurable as profit and send it to engineering. 
After all, high-cost laboratory-based subjects receive more funding per output than do theoretical ones.
However, a world-leading or excellent  REF case study delivers infinitely more 
than less recognised  ones.
Any move to submit to engineering  would collapse statistical physics in Coventry and hence collapse the rich and rewarding collaborations with Lviv and other partners --- collaborations that have impacted the world.

\section{Suggestions}
\label{6}

``If it wasn't for academic freedom, you wouldn't be working on this stuff.''
Words to that effect came from a previous Head of Department when I first started my adventures in the scientometrics world~\cite{1=44=EPL}.
Ventures into  interdisciplinarity  are  risky --- you have to be bold enough to step beyond the comfort zone of statistical physics and academic freedom liberates {you} to do just that.

To my mind, at least this is precisely what REF has in mind and what the impact agenda encourages.
The  Stern Report calls for ``widening and deepening the notion of impact to include influence on public engagement, culture and teaching as well as policy and applications more generally. It also suggests to ``tackle the underrepresentation of interdisciplinary research in the REF''. 

The Metrics Tide and Stern reports helped save our world from wonton and misplaced usage of metrics by maintaining  REF  in the hands of academics  (peer reviewers).
They saved the nation from ``metricide''.
However, as we have seen, there is a difference between  extensive and intensive, and  the impact agenda has opened a new can of worms at local levels in some universities. 
It has spawned the rise of a new professional class who interpret things their way, and that way can be at odds with the best of intentions of REF.

I know of a number of instances where academics shy away from declaring or developing their impact because of the perils of  contact with professional services. 
They fear being drawn away from gaining the only recognition they really value --- recognition by peers in their own academic domains.
This means that impact potentially generated may not be developed.
This discouragement is partly a legacy issue. I have heard many accounts of academics being hired, fired, promoted and demoted on the basis of metrics and single-opinion peer reviews in ``mock-REF''  exercises.

The intent of REF is laudable in my opinion. However, as with prisoners dilemma, games are played and, in the ultra-competitive world of individualised metrics and targets, absence of trust leads to a clash between short-term individual interest  and long-term collective goals.
Consider the case of a senior academic supposedly versed in the laws of impact and with a case study they have invested time developing. 
At the last minute along comes a more junior researcher, not versed in the rules of REF, with a single output but irrefutably world-leading (4*) impact. 
Suppose submission of the latter as an impact case study would dislodge the former.
What then does the senior do? 
A good person will stand aside and let the junior through. 
But, as we have seen, not all academics are good and a selfish one may pretend {or even believe} that one output is indeed not enough for a case study.
Alice in wonderland  believes withholding information is the same as lying and I 
agree~\cite{Alice}.
With the increased importance of impact for funding, having a REF impact case study based on your research could be a career maker/breaker in our very competitive environment; not everyone would do what Alice would.

For REF2021, ``the new requirement to return all staff who have a significant responsibility for research is about promoting inclusion, about addressing the negative consequences for staff associated with exclusion in previous exercises, and about encouraging the presentation of a rounder picture of research activity. But it is also about underlining the move away from the focus on individuals, and their related publications and other outputs''.

I suggest this very laudable motive could be more effectively realised in the future by further loosening of the laws of REF --- submit more papers and more impact case studies than needed. 
As an example, suppose up to 4 outputs per head could be submitted but only the top 2.5 papers per head would count. 
For 20 staff that would mean liberty to submit up to 80 papers of which only the top 50 count for the aggregate REF score.
That way nobody in the submitting institution would know which top 50 these are --- competition is removed, collegiality restored and no individual is hired or fired on the basis of variable or metricised mock-REF scores. 
Sure, that will require more assessors but help is to hand --- use Calibrate with Confidence! 
Multiple tests have shown  that the algorithm is robust to diluting the number of assessors assigned to each object assessed.

For impact case studies I suggest similar and more. 
I suggest to loosen the restriction of 5 pages. 
Our MMM case study is multidimensional and has enabled, enriched, influenced, informed and changed the performance, policies, practices, products, services, understanding, awareness and well-being of a very diverse set of beneficiaries. This has been achieved through media, public engagement, industrial cooperation and has inspired community events, performances, art, poetry,  theatre, etc. It is direct and indirect and its sheer vastness worries me that the wood will be lost for the trees. 
To avoid having the cram it all into a one-size-fits-all document I suggest more liberty.
So how many pages? 
Well {\ldots}if you are still reading this now you have made your way through 17 pages --- you decide.

So there you have it --- some of my thoughts in the run-up to REF2021.
In constructing this paper I drew on some of my own experiences and experiences of colleagues in various universities. 
I used some (but not all) of our scientometrics-based impact to tell this story but also a little from  
{\emph{Maths Meets Myths}}.
I hope to  relate the latter story on another occasion.

{One version of a famous quote attributed to many sources is: ``measure what you value because you will value what you measure''.
My ``take-away message'' is let academics measure academia, including impact emanating from it, because it is we who value it most.}
Let's loosen the rules so that REFs (in the hands of calibrated academics)  and not mock-REFs (in the hands of {uncalibrated} professional services) determine destinies.
Professional services can help generate impact and gather evidence of it but overly linear and austere views should not determine who, what or where to submit.

\section{Acknowledgements}
\label{7}

This paper  came about through years of \emph{trusted} collaboration.
I don't list all I should thank because of space limitations but they know who they are and I will thank them in person.
But here I would especially like to mention some who were involved in this particular paper.
The research underpinning most of it was carried out with 
Bertrand Berche, Yurij Holovatch, Olesya Mryglod, Rob Low, Sarah Parker and Robert MacKay.
I also thank Martine Barons, Colm Connaughton, Rosemary Harris, Biagio Lucini, Elaine Crooks, Eugene Lytvynov for points of view from academic impact leads, research heads and reviewers at other Universities.
I thank Thalia Cooley and Laura Forster for students' perspectives; 
Andrea Crowley, Tom Evans, Elena Gaura, Don Harris and Laura Noble for countless discussions which ``enriched my deliberations and provoked me to ask questions that might otherwise remain unaddressed'';
Steve Edwards for elucidating the meaning of ``professional'';
Tim Ellis for many productive discussions on science and philosophy;
Reinhard Folk for physics' perspectives;
Neil Forbes for sharing the word ``metricide'';
Charo del Genio for the term ``corporate overlords'';
Susanne Horn, Thierry Platini and Taras Yavorskyi for badly needed sanity checks;
Madeleine Janickyj for having what it takes to carry the flag;
%Annette Kinsella for engaging and being engaging;
Abhishek Kumar spreading the word;
Rob Low for knowing everything about mathematics and physics and willingness to share;
Marisa McGlinchey for inspiration on holding the anti-ferromagnetic line in a ferromagnetic world;
Alban Potherat for being more a colleague than a line manager;
Robin de Regt and Joseph Yose for sympathetic and valuable insights from the academic-corporate borderland;
Coventry University for accepting academic freedom and taking risks for 18 years.
Finally I thank Camille No\^{u}s,  the virtual scientific personality who is the antithesis of vile plagiarists. 
Camille represents how it should be --- ``the collective and open character of the creation and dissemination of knowledge, under the control of the academic community''~\cite{Camille}.

\ukrainianpart
\title[Two Cultures]%
{Дві культури: ``Вони і ми'' в академічному світі}
\author[R. Kenna]{Р. Кенна\refaddr{label4,label3}}
\addresses{
	\addr{label4}Група статистичної фізики, Центр прикладних математичних дослiджень, Унiверситет Ковентрi, Ковентрi, Англiя
	Centre for Fluid and Complex Systems, Coventry
 \addr{label3}Докторський коледж статистичної фiзики складних систем, Ляйпцiґ-Лотарингiя-Львiв-Ковентрi  $({\mathbb L}^4)$}
\makeukrtitle 
	\begin{abstract}
\tolerance=3000%
{\emph{Вплив}} академічних досліджень на неакадемічний світ стає все більш важливим в той час, як органи влади прагнуть домогтися {підзвітності} за державні  інвестиції. Цей вплив відкриває нові можливості для {так званих} ``менеджерських служб''\,:  оскільки наукометричні інструменти обдаровують декого впевненістю, що вони можуть кількісно визначити якість, а порядок денний такого впливу дає змогу  вимірюванням  оцінити дане дослідження.
Ця стаття була частково навіяна відомою дискусією про ``дві культури'', спровокованою Ч.П. Сноу понад 60 років тому. Він побачив провалля між різними академічними дисциплінами,  а я бачу провалля між науковцями і менеджерськими службами, які пов'язані лише контактами через конкуруючі цілі.
Ця стаття описує  мій власний досвід та досвід, переказаний моїми колегами з різних університетів.
Метою статті є започаткування дискусії серед осіб,  зацікавлених в покращенні академічного середовища
у дусі співпраці та конструктивізму.
Як сказав один з колег з менеджерської служби, те, що я маю сказати, ``мусить бути сказано''.
Я маю велику приємність подати цю статтю до збірки праць, що присвячена  60-літтю відомого фізика, мого доброго приятеля та колеги Ігора Мриглода.
Роль Ігора як лідера Інституту фізики конденсованих систем у Львові була важливою для створення  впливу, описаного в цій роботі, і є ключовим елементом розповіді, яку я хочу вам розповісти. 
\keywords  дві культури, оцінювання досліджень, REF, вплив, наукометрія, термодинаміка, статистична фізика.
\end{abstract}

\end{document}